\documentclass[aps,prd,floatfix,preprintnumbers,letterpaper,ifmbe,twocolumn,usegraphicx]{revtex4}
\usepackage{graphicx}
\usepackage{graphics}
\usepackage{epsfig}
\usepackage{amsfonts,amsmath,amsfonts,amscd}

\begin{document}

\title{Black hole in a superconducting plasma}

\author{Chinmoy Bhattacharjee$^{\,1}$, Justin C. Feng$^{\,2,3}$,
S.M. Mahajan$^{\,4,5}$}

\affiliation{$^{1}$Department of Physics, Rutgers University Newark, Newark 07102 USA}
\affiliation{$^{2}$St. Edwards University, Austin, TX 78704 USA }
\affiliation{$^{3}$Theory Group, Department of Physics,The University of
Texas at Austin, Austin, TX 78712 USA}
\affiliation{$^{4}$Department of Physics,The University of
Texas at Austin, Austin, TX 78712 USA}
\affiliation{$^{5}$SN University, Uttar Pradesh, India}

\begin{abstract}
The generalized vortical formalism provides an electrodynamic description for superconducting states---in the generalized vortical formalism, a superconducting state may be defined by the vanishing of an appropriate generalized vorticity and characterized by zero generalized helicity for incompressible fluids. In this article, we investigate these states for compressible plasmas in black hole spacetime geometries using the curved spacetime generalization of the grand generalized vortical formalism. If the magnetic field is axisymmetric and the thermodynamic properties are symmetric about the equatorial plane, the resulting states are characterized by a vanishing skin depth and a complete expulsion of the magnetic field at the equator of the black hole horizon. Moreover, if the thermodynamic properties of the plasma are uniform at the horizon, we find that the magnetic field is completely expelled from the horizon, and the plasma behaves as a perfect superconductor near the horizon. This result is independent of the spin of a black hole, holding even for a (nonrotating) Schwarzschild black hole, and demonstrates that the geometry near black hole horizons can have a significant effect on the electrodynamics of surrounding plasmas.
\end{abstract}

\maketitle

\section{\label{sec:level1}Introduction}

In addition to infinite conductivity, the defining feature of a conventional superconductor (CS) is the expulsion of magnetic flux from its interior---the Meissner effect. The phenomenological London equation 
\begin{equation}\label{londonequation}
    \nabla^2\vec{B}=\frac{\vec{B}}{
    \lambda_s^2},
\end{equation}
expresses the fact that the magnetic field decays (from its external value) in a skin depth, $\lambda_s=c/\omega_p$ ($\omega_p=\sqrt{4\pi n q^2/m}$ is the plasma frequency), a length scale characteristic of the charge carriers in the superconducting `fluid'. For a canonical superconductor, this superfluid consists of Cooper pairs formed through quantum interactions. 

What is rather interesting is that this unique property associated with conventional superconductivity is electro-dynamically equivalent to the complete elimination of a `generalized vorticity' (GV) in the representative (incompressible) fluid \cite{mahajan2008classical}, that is
\begin{equation}
\nabla\times\vec{V}+ \frac{q \vec{B}}{M}=0,
\end{equation}
where the first term denotes the standard fluid vorticity (rotation of the fluid velocity $\vec{V}$) while the second term, proportional to the cyclotron frequency, could be viewed as the `vorticity' of the electromagnetic field. One may write the above equation in the equivalent form,
\begin{equation}\label{Vort}
\vec{\Omega} = \vec{B} + \frac {M}{q}\nabla\times\vec{V}=0,
\end{equation}
and $\vec{\Omega}$ will be the generic symbol for GV. For the CS for which the dynamic fluid is a collection of Cooper pairs, $M=2m$ and $q=-2e$. When coupled with Ampere's law $ \nabla\times\vec{B}= 4\pi n(-2e)\vec{V}$, Eq.~(\ref{Vort}) reproduces the correct skin depth. 

It was noticed in \cite{mahajan2008classical} that the electrodynamics (leading to the London equation) is blind to the microscopic origin of Eq.~(\ref{Vort})---it could be quantum or non-quantum. In this sense, the CS is a special case of the superconducting state where quantum correlations led to the specific $\vec{\Omega}_{CS}=0$. More directly stated, the behavior of the magnetic field in a CS is entirely equivalent to the vanishing of the canonical vorticity everywhere, including the skin depth region where the magnetic field is zero. Thus, independent of the origin of superconductivity, the electrodynamic signature of a CS is the vanishing of the canonical vorticity of the appropriate superconducting fluid. One could attempt a broad generalization by elevating a condition of the type epitomized in Eq.~(\ref{Vort})---{\it the vanishing of an appropriate generalized vorticity}---to constitute the very definition of a superconducting state. 

In curved spacetime, the expulsion of magnetic fields is not unique to superconducting media. Under a certain set of assumptions, a spinning black hole will expel magnetic fields near the horizon in the extremal limit. A rather general argument by Penna \cite{penna2014black} shows that a black hole will expel axisymmetric magnetic fields as its spin approaches the extremal limit (provided that split-monopole type solutions are excluded), and it has been shown explicitly in the context of Wald's solution \cite{Wald1974} for magnetic (vacuum test) fields that the magnetic field does indeed vanish at the horizon of a spinning black hole in the extremal limit \cite{Kingetal1975,BicakDvorak1980,Chamblinetal1998,komissarov2007meissner}. It was later shown that the expulsion of magnetic fields from the horizon still occurs for magnetic fields generated by plasmas surrounding the black hole \cite{bivcak2015near,gurlebeck2017meissner,gurlebeck2018meissner,kunz2017magnetized}. The expulsion of magnetic fields from horizons in the extremal limit suggests that the extreme conditions near black hole horizons can have some profound effects on the electrodynamical properties of plasmas surrounding black holes.

This superconducting-like phenomena near extremal horizons motivates further investigation of the effect of horizon geometry on the electrodynamical properties of magnetized plasmas. In particular, we exploit the electrodynamical similarity between a canonical superconductor and a vorticity-free hot relativistic fluid (which we call a superconducting plasma) to further explore the properties of plasmas surrounding black holes (we do not assume extremal or near extremal spin) that exhibit superconducting-like behavior. This requires extending the definition of classical generalized vorticity to relativistic plasmas. The relativistic formalism, termed `magnetofluid unification' (henceforth the magnetofluid formalism) and first proposed in (\cite{PhysRevLett.90.035001}) and later extended in \cite{asenjo2013generating,bhattacharjee2015magnetofluid}, replaces the regular fluid vorticity $(\vec{\nabla}\times\vec{V})$ with a composite thermo-kinetic fluid vorticity $(\vec{\nabla}\times\mathcal{G}\Gamma\vec{V})$ where $\mathcal{G}$ is the thermodynamic factor and $\Gamma$ is the Lorentz factor.

Recently, the magnetofluid formalism, which seemed to valorize the magnetic part of the electromagnetic part, was renamed Electro-Vortic (EV) formalism \cite{mahajan2016relativistic}, to explicitly emphasize its covariant character. It was also shown that for a restricted form of the velocity field (the Clebsch form, $ U^\mu = (\nabla^\mu \mathcal{Q})/T$), a generalized helicity is absolutely conserved irrespective of the thermodynamics governing a perfect relativistic fluid. The appropriate vorticity (the magnetic part of the EV field), termed the grand generalized vorticity, takes the form
\begin{equation}
\vec{\Omega}_G=\vec{B}+m/q \vec{\nabla}\times\mathcal{G}^\prime T\vec{U} 
\end{equation}
where
\begin{equation}
    \mathcal{G}^\prime=\left(\frac{\mathcal{G}}{T}-\frac{\sigma}{m}\right).
\end{equation}
and depends on the entropy density $\sigma$, and flow velocity is represented as $\vec{U}=\Gamma\vec{V}$. In this letter, we investigate the characteristics of plasma electrodynamics due to the complete expulsion of the grand generalized vorticity $\Omega_G=0$ (instead of the generalized vorticity $\Omega=0$) near a black hole---such equilibrium states, will perhaps, constitute a more complete description of a superconducting plasma (as defined by its macroscopic electrodynamical properties) surrounding a black hole.

We begin with a brief overview of the standard magnetofluid formalism for plasma dynamics in curved spacetime. Next, we discuss the grand generalized vortical formalism in curved spacetime and we obtain the magnetic field profile for the stationary $\Omega_G=0$ solution in a stationary, axially symmetric spacetime. Finally, we perform a skin depth analysis for the superconducting $\Omega_G=0$ states, and discuss the features of these plasma states in the vicinity of the horizon of black holes.

\section{\label{sec:level2}Plasma dynamics in Curved spacetime- Grand Generalized Vorticity}

Here, we describe the essential elements of the EV formalism in curved spacetime [\cite{PhysRevLett.90.035001}, \cite{bhattacharjee2015magnetofluid,bhattacharjee2018surveying}, \cite{mahajan2016relativistic}]. A multi-species plasma in curved spacetime is governed by the following equation of motion
\begin{equation}
\label{eomfluid}
mnU^{\nu}\nabla_{\nu}\left(\mathcal{G}U^{\mu}\right)=qnF^{\mu\beta}U_{\beta} - \nabla^{\nu}p ,
\end{equation}
where the quantities $m$ and $q$ are the respective mass and charge of the constituent particles of the fluid, while $n$ and $p$ respectively denote the number density and the pressure. The fluid four-velocity for each species may be written as $U^{\mu}={dx^{\mu}}/{d\tau}$, where $\tau$ is the proper time for a fluid element. The thermodynamic factor $\mathcal{G}$ is given by the expression $(\rho+p)=h=mn\mathcal{G}$ with $h$ and $\rho$ being the respective enthalpy and mass density of the fluid.

The defining step in the initial EV formalism is the ``construction" of a new temperature-transformed flow field tensor \cite{bekenstein1987helicity,PhysRevLett.90.035001}  $S^{\mu \nu}:= \nabla ^{\mu}\left(\mathcal{G}U^{\nu}\right)-\nabla^{\nu}\left(\mathcal{G}U^{\mu}\right)$, in terms of which, Eq.~(\ref{eomfluid}) can be rewritten as
\begin{equation}
\label{eommagnetofluid}
    qU_{\mu}\mathcal{M}^{\mu\nu}=T\nabla^{\nu}\sigma,
\end{equation}
where $\mathcal{M}^{\mu\nu}=F^{\mu\nu}+\left(m/q\right)S^{\mu\nu}$ is the magnetofluid tensor, and the entropy density $\sigma$ for the fluid obeys $T\nabla^{\nu}\sigma = \left(mn\nabla^{\nu}\mathcal{G}-\nabla^{\nu}p\right)/n$. One readily notices that a relativistic perfect fluid is isentropic,
\begin{equation}
\label{isentropic}
    U_{\nu}\nabla^{\nu}\sigma=0,
\end{equation}
the entropy density $\sigma$ being constant along a flow line. 

Equation (\ref{eommagnetofluid}) for plasma dynamics can be recast in a source free form by defining the following EV potential $\mathbb{P}^\mu$:
\begin{equation}\label{GGVpotential}
    \mathbb{P}^{\mu}=A^{\mu}+\frac{m}{q} \, \mathcal{G} \, U^{\mu} - \sigma \, \nabla^\mu \mathcal{Q} ,
\end{equation}
where $A^{\mu}$ and $\mathcal{Q}$ are four potential and a scalar, respectively. Then, the covariant equation of motion Eq.(\ref{eommagnetofluid}) may be written: 
\begin{equation}\label{grandequation}
    qU_{\mu}\mathbb{M}^{\mu\nu}=0,
\end{equation}
where 
\begin{equation}
    \mathbb{M}^{\mu\nu}=\nabla^{\mu}\mathbb{P}^{\nu}-\nabla^{\nu}\mathbb{P}^{\mu}.
\end{equation}
The scalar $\mathcal{Q}$ must satisfy the following:
\begin{equation}\label{ZandTemperature}
U_\nu \nabla^\nu \mathcal{Q} = T / q 
\end{equation}
so that one may recover Eq.(\ref{eommagnetofluid}) from Eq.(\ref{grandequation}).
In general, flow fields $U^\mu$ satisfying Eq.~(\ref{ZandTemperature}) can be written in the form
\begin{equation} \label{SCSC-Clebsch}
T U^\mu = \nabla^\mu \mathcal{Q} + b^\mu,
\end{equation}
where the vector $b^\mu$ is orthogonal to $U^\mu$. To simplify the analysis, we assume $b^\mu=0$, so that $T U^\mu = \nabla^\mu \mathcal{Q}$. \footnote{One might worry that this implies that the flow field $U^\mu$ is irrotational, which may be the case if one insists that Eq. (\ref{SCSC-Clebsch}) is a global condition for the flow field. However, if this is a local condition, and if the temperature is not constant, the expression $U^i=(1/T)\nabla^i \mathcal{Q}$ has nonvanishing curl, and forms part of the locally valid Clebsch decomposition for a vector field \cite{wu2007vorticity}. In fact, one can go further, and obtain an almost-global (in the sense that it is valid for all but a measure zero set of the manifold) description for the flow field from a potential in appropriate coordinates---from the converse of Poincar{'e}'s lemma for one-forms one can show that this is possible if the domain of the vector field is contractible in some coordinates. For example, one can construct a rotating vector field in polar coordinates $r,\phi$ (in Euclidean space) from the potential function $\Phi(r,\phi)=\phi$.}

We briefly describe the application of the $3+1$ (ADM) decomposition of spacetime to the EV formalism, which is more thoroughly discussed in \cite{bhattacharjee2015magnetofluid,bhattacharjee2018surveying}. In the $3+1$ (ADM) formalism, spacetime is foliated by a family of 3-dimensional spacelike hypersurfaces such that each hypersurface $\Sigma_t$ is defined by a constant value for the time coordinate $t$. The line element in the ADM formalism takes the form
\begin{equation}\label{canonicalmetric}
	ds^2=-\alpha^2dt^2+\gamma_{ij}(dx^i+\beta^i dt)(dx^j+\beta^j dt)
	\end{equation}
where $\alpha$ is the lapse function, $\beta^i$ is the shift vector, and $\gamma_{ij}$ is the spatial metric (the induced metric) for the hypersurface $\Sigma_t$; collectively, $\alpha$, $\beta^i$ and $\gamma_{ij}$ form the ADM variables. Equation (\ref{grandequation}) is split into space and time components; the spatial components, which form the 3D equation of motion, are obtained by applying the projection operator $\gamma{^\mu}{_\nu}=\delta^{\mu}\ _{\nu}+n^{\mu}n_{\nu}$ (where $n^\mu=-\alpha g^{\mu \nu} \nabla_\nu t$ is the unit normal vector to $\Sigma_t$):
\begin{equation} \label{SCSC-EoMgen1}
\alpha\Gamma \vec{\mathcal{E}}_G + \alpha\vec{U} \times \vec{\Omega}_G = 0.
\end{equation}
Here, the grand generalized electric and vorticity are given by the respective equations:
\begin{align}\label{generalefield}
	\vec{\mathcal{E}}_G= \ & \vec{E}-\frac{m}{\alpha q}\vec{\nabla}(\alpha T \mathcal{G}^\prime\Gamma)
	-\frac{m}{q}\left[2\underline{\underline{{\sigma}}}\cdot(\mathcal{G}^\prime T \vec{U})+\frac{2}{3} K \mathcal{G}^\prime T \vec{U}\right] \notag\\
\ &-\frac{m}{q\alpha}\left(\partial_t(\mathcal{G}^\prime T \vec{U}) -\mathcal{L}_{\vec{\beta}}(\mathcal{G}^\prime T  \vec{U})\right);
	\end{align}
\begin{equation} \label{SCSC-GGV}
   \vec \Omega_{G}=\vec{\nabla}\times \vec{\mathbb{P}}_{G}=\vec{\nabla}\times\left(q \vec{A} + m \mathcal{G}^\prime T \vec{U}\right),
\end{equation}
where $\Gamma=1/\sqrt{1-V^2}$ is the Lorentz factor and $\vec A$ is the vector potential. Here, $\vec{\beta}$ is the shift vector, $\mathcal{L}_{\vec{\beta}}$ is the Lie derivative with respect to $\vec{\beta}$, $\underline{\underline{{\sigma}}}$ is a trace-free rank-2 tensor (with components $ \sigma{^i}{_j}$ formed from $\partial_t{\gamma}_{ij}$ and $\beta^i$) called the shear tensor, and $K$ is the mean curvature for $\Sigma_t$. The dynamics of a hot, relativistic, magnetized plasma as expressed by Eq.~(\ref{SCSC-EoMgen1}) has a strikingly similarity with ideal MHD model i.e. it has the structure of an ideal Ohm's law.

\section{The $\Omega_G=0$ solution}
Eq.~(\ref{SCSC-EoMgen1}) is trivially satisfied if the grand generalized electric field $\vec{\mathcal{E}_G}$ and grand generalized vorticity $\vec{\Omega}_G$ (GGV) are, separately, zero
\begin{align}\label{Superstate} 
     \vec{\mathcal{E}}_G & =\nabla(\alpha T \Gamma\mathcal{G}^\prime)-\alpha\left[2\underline{\underline{{\sigma}}}\cdot(\mathcal{G}^\prime T \vec{U})+\frac{2}{3} K \mathcal{G}^\prime T \vec{U}\right] \nonumber \\
    &\qquad +\mathcal{L}_{\vec{\beta}}(\mathcal{G}^\prime T\vec{U}) =0 
     \end{align}
     \begin{align}\label{superstate 2}
     \vec{\Omega}_G=q \, \vec{B} + m \, T \, \vec{\nabla} \mathcal{G}^\prime \times \vec{U}=0,
\end{align}
where magnetic field is $\vec{B}=\vec{\nabla}\times\vec{A}$ and we have made use of the vector identity $\vec{\nabla} \times \vec{\nabla} \mathcal{Q} = 0$. The first equation is a general relativistic Bernoulli's condition, and is not essential to understanding the $\Omega_G=0$ state under investigation. The second equation describes the total expulsion of GGV and is, clearly, a generalization of London equation (expulsion of the standard canonical vorticity). A state with zero GGV must, necessarily, have a zero grand generalized helicity (GGH). It should be emphasized here that $\vec{\Omega}_G=0$  (along with the Bernoulli  condition)  is an exact  solution to the time-independent GGV evolution equation\cite{bhattacharjee2018surveying}.

Since we are solving steady-state charge neutral fluid-Maxwell system, Eq. (\ref{superstate 2}) should be coupled with the steady state Ampere's law
\begin{equation}\label{Amph}
    \vec{\nabla}\times(\alpha \vec{B})=4\pi nq\alpha\vec{U},
\end{equation}
when normalized in terms of the cyclotron frequency, $B_c=q/m \, B$, becomes 
\begin{equation}\label{Amphmod}
    \vec{\nabla}\times({\alpha \vec{B}_c})=\frac{\hat{n}}{\lambda^2}\alpha \vec{U},
\end{equation}
where $n= \hat{n} n_0$ , $\hat{n}$ is the density envelope, and the skin depth $\lambda = \sqrt{4 \pi n_0 q^2/m}$, associated with some average density, is an intrinsic length scale of the dynamics. 
Using Eq.~(\ref{Amphmod}), we rewrite Eq.~(\ref{superstate 2}) as
\begin{equation}\label{SCSC-OmegaZeroState3}
\vec{\mathcal{B}} + \frac{\lambda^2 T}{\hat{n}} \left[ \nabla \mathcal{G}^\prime \times (\nabla \times \vec{\mathcal{B}}) \right] = 0,
\end{equation}
where we have defined $\vec{\mathcal{B}}=\alpha \vec{B}_c$. Notice that despite its somewhat complicated form, Eq.~(\ref{SCSC-OmegaZeroState3}) is linear and a generalization of the London equation in a hot relativistic fluid signifying the complete expulsion of the grand canonical vorticity.

We now seek a simple  axisymmetric solution of Eq.(\ref{SCSC-OmegaZeroState3}). The general line element for a stationary and axisymmetric spacetime is \cite{Wald,ChoquetBruhat2015}:
\begin{equation}\label{SCSC-StationaryLineElement}
ds^2 = - \tilde{\alpha}^2  dt^2 + 2 \, \beta_\theta \, dt \, d\phi + h^2_1 \, dr^2 + h^2_2 \, d\theta^2 + h^2_3  \, d\phi^2 ,
\end{equation}
 where the quantities $\tilde{\alpha}^2$, $\beta_\theta$, $h_1$, $h_2$, and $h_3$ are all assumed to be functions of $r$ and $\theta$ only. The quantity $\tilde{\alpha}=\sqrt{|g_{tt}|}$ is not the ADM lapse function, but is related to the ADM lapse function $\alpha$ through the expression:
$g_{tt}=-\tilde{\alpha}^2=-\alpha^2+\gamma_{ij} \beta^i \beta^j$. Here, $\gamma_{ij}$ is the induced metric and $\beta^i$ is the shift vector and the shift vector corresponds to the $\theta$ component of $\beta_i=\gamma_{ij} \beta^j$.

Assuming thermodynamic quantities (in particular $\mathcal{G}^\prime$) depend on $r$ and $\theta$ direction, Equation (\ref{SCSC-OmegaZeroState3}) can be written as
\begin{equation}\label{SCSC-EqBr}
B_r - \frac{\zeta \mathcal{G}^\prime_{,\theta}}{\alpha h_1 h^2_2}  \left( \frac{\partial (\alpha h_1 B_r)}{\partial \theta}-\frac{\partial (\alpha h_2 B_\theta)}{\partial r}\right) = 0
\end{equation}
\begin{equation}\label{SCSC-EqBtheta}
B_\theta + \frac{\zeta \mathcal{G}^\prime_{,r}}{\alpha h_2 h^2_1} \left( \frac{\partial (\alpha h_1 B_r)}{\partial \theta} - \frac{\partial (\alpha h_2 B_\theta)}{\partial r}\right) = 0
\end{equation}
\begin{equation}\label{SCSC-EqBphi}
B_\phi  - \frac{\zeta}{\alpha h_3} \left(  \frac{\mathcal{G}^\prime_{,\theta}}{h^2_2} \frac{\partial (\alpha h_3 B_\phi)}{\partial \theta} +  \frac{\mathcal{G}^\prime_{,r}}{h^2_1} \frac{\partial (\alpha h_3 B_\phi)}{\partial r}\right) = 0
\end{equation}
where $B_r$, $B_\theta$, and $B_\phi$ are the orthonormal basis components of $\vec{B}_c$, and we have defined $\zeta=\lambda^2 T/\hat{n}$, $\mathcal{G}^\prime_{,\theta}=\partial \mathcal{G}^\prime/\partial \theta$ and $\mathcal{G}^\prime_{,r}=\partial \mathcal{G}^\prime/\partial r$.

If one assumes that the thermodynamic properties of the fluid are symmetric about the equatorial plane, then $\mathcal{G}^\prime$ has a maxima or minima in $\theta$ at the value $\theta=\pi/2$. One may conclude that at the equatorial plane $\theta=\pi/2$, $\mathcal{G}^\prime_{,\theta}=0$. Immediately, we see that inserting the condition $\mathcal{G}^\prime_{,\theta}=0$ into Eq. (\ref{SCSC-EqBr}) implies that the radial component $\mathcal{B}_r$ of the magnetic field vanishes; the radial component of the magnetic field $\mathcal{B}_r$ therefore vanishes at the equatorial plane $\theta=\pi/2$.
The solutions in the equatorial plane $\theta=\pi/2$ for the remaining components of the magnetic field are then

\begin{equation}\label{SCSC-OmegaZeroStateSolnTh}
B_{\theta} = B_{\theta,0}  \exp \left(\int^\infty_r \frac{dr^\prime}{Z(r^\prime)}\right)
\end{equation}
\begin{equation}\label{SCSC-OmegaZeroStateSolnPh}
B_{\phi} = B_{\phi,0} \exp \left(\int^\infty_r \frac{dr^\prime}{Z(r^\prime)}\right).
\end{equation}
where$Z(r) := \left.\zeta \mathcal{G}^\prime_{,r}  \right|_{\theta=\pi/2}$, and $B_{\theta,0}$ and $B_{\phi,0}$ are constants determined by boundary conditions.

For the Schwarzschild geometry, $h_2=r\sin \theta$ and $h_3=r$, which indicates the magnetic field profiles decay as $r\rightarrow \infty$.  The exact profiles can be determined by computing the form of $Z(r)$ which will depend on some equation of state and the thermodynamic properties of the plasma. This will be explored further in future work.

\subsection{Magnetic Field Expulsion and Skin Depth Analysis}
We have seen that the symmetry of $\mathcal{G}^\prime$ about the equatorial plane implies the vanishing of $B_r$ at the equatorial plane. We will now examine the behavior of magnetic fields, as described by Eqs.~(\ref{SCSC-EqBr}-\ref{SCSC-EqBphi}), at the horizon of an axisymmetric black hole. For the Schwarzschild spacetime, $h_1^{-2}=\alpha^2=1-2 M/r$, which vanishes at the horizon surface defined by $r=2 M$ (also note that $\alpha h_1=1$). In the Kerr spacetime, one also finds that $h_1^{-2}=0$ at the horizon and $\alpha h_1$ remains finite. From Eq. (\ref{SCSC-EqBtheta}), we see that at the horizon, the $\theta$-component of the magnetic field is completely expelled from the horizon, or that $B_\theta=0$; we emphasize here that $B_\theta$ vanishes on the \textit{entire} horizon. Under the condition $\mathcal{G}^\prime_{,\theta}=0$, which we assume is valid at the equatorial plane, we find [from Eqs. (\ref{SCSC-EqBr}) and (\ref{SCSC-EqBphi})] that the remaining components of the magnetic field vanish at the horizon. One may therefore conclude that at the equator of the black hole horizon (where we expect $\mathcal{G}^\prime_{,\theta}=0$), all components of the magnetic field vanish; the magnetic field is completely expelled from the equator of a black hole horizon. The plasma therefore behaves as a perfect superconductor at the equator of a black hole horizon. We observe that the symmetry in $\mathcal{G}^\prime$ about the equatorial plane need not be exact; the complete expulsion of magnetic fields from the horizon occurs for any value of $\theta$ for which $\mathcal{G}^\prime$ has a minimum or a maximum. Moreover, if the thermodynamic properties of the plasma are uniform near the horizon, so that $\mathcal{G}^\prime$ is constant, the magnetic field is completely expelled from the horizon, and the entire black hole is immersed in a plasma which behaves as a perfect superconductor at the horizon.

To further establish this result, it is appropriate to work in terms of the proper distance $dR=h_1 dr$. Again, we examine the properties of the plasma at the equator, so that $\mathcal{G}^\prime_{,\theta}=0$. When written in terms of $B_{\theta}$ and $B_{\phi}$ components, Eqs.~(\ref{SCSC-EqBtheta}) and (\ref{SCSC-EqBphi}) can be rearranged into a slightly different form
\begin{align}\label{skindepth1}
\frac{1}{\tilde{\lambda}^2} & = \frac{\partial \ln \mathcal{G}^\prime}{\partial R} \,\frac{\partial }{\partial R} \ln[ h_2 \, B_\theta\alpha ]\nonumber \\
& = \frac{\partial \ln \mathcal{G}^\prime}{\partial R} \left[\frac{\partial \ln (h_2 \, B_\theta)}{\partial R} + \frac{\partial \ln \alpha}{\partial R} \right]
\end{align}
\begin{align}\label{skindepth2}
\frac{1}{\tilde{\lambda}^2} &= \frac{\partial \ln \mathcal{G}^\prime}{\partial R} \,\frac{\partial }{\partial R} \ln [ h_3 \, B_\phi \alpha]\nonumber \\
& = \frac{\partial \ln \mathcal{G}^\prime}{\partial R} \left[\frac{\partial \ln (h_3 \, B_\phi)}{\partial R} + \frac{\partial \ln \alpha}{\partial R} \right],
\end{align}
where $\tilde{\lambda}^2=\zeta\mathcal{G}^\prime$ is the modified skin depth characterizing the $\Omega_G=0$ solution. Eqs.(\ref{skindepth1}-\ref{skindepth2}) express the skin depth of the plasma in terms of variations of magnetic field and thermodynamic gradients whereas, for classical superconducting state, the skin depth depends only on the variation of the magnetic field.
In addition to magnetic and thermodynamic variations, the term $\partial \ln \alpha/\partial R$ is purely a general relativistic correction to the skin depth and can be computed as follows
\begin{equation}\label{SCSC-AlphaDeriv}
\frac{\partial \ln \alpha}{\partial R} = \frac{1}{\alpha} \frac{\partial \alpha}{\partial R} = \frac{1}{h_1 \alpha} \frac{\partial \alpha}{\partial r} 
\end{equation}
For Schwarzschild black hole, Eq.~(\ref{SCSC-AlphaDeriv}) becomes
\begin{equation}
\frac{\partial \ln \alpha}{\partial R} = \frac{G M}{r^2 \sqrt{1-2 G M/r}},
\end{equation}
which diverges at the horizon---in the limit $r \rightarrow 2 G M$. One can show that the divergent behavior at the horizon also persists  for rotating black holes (even for the non-extremal case). It follows that the skin depth vanishes at the horizon, and the plasma behaves as a perfect superconductor at the equator $\theta=\pi/2$ of the horizon if $\mathcal{G}^\prime$ is symmetric about the equator, or on the entire horizon if $\mathcal{G}^\prime$ is uniform. The skin depth analysis demonstrates that this effect is coordinate independent, as the analysis is carried out in terms of proper length, and also that this effect is a consequence of the spacetime geometry near the horizon of a stationary black hole. Note also that the complete expulsion of magnetic fields at black hole horizon or horizon equator is independent of the value of the spin, in contrast to the case with vacuum magnetic test fields, which are only expelled from the horizon in the limit of extremal spin.

\section{conclusion}
By defining a superconducting plasma as the one in which the grand generalized vorticity (GGV) vanishes identically ($\Omega_G=0$), we have examined the magnetic properties of such a state for a magnetized plasma around a black hole. The principal features of this state (applicable to both rotating and nonrotating black holes) may be summarized as follows: (i) the grand generalized helicity associated with the vortical field lines is identically zero (ii) the $\theta$-component $B_\theta$ of the magnetic field vanishes at the horizon regardless of the thermodynamic properties of the plasma (this is purely due to near-horizon geometry), (iii) assuming symmetry in $\mathcal{G}^\prime$ about the equatorial plane the magnetic fields are expelled according to a varying scale length which becomes zero at the equator of the black hole horizon---no restrictions are placed on laminar flows, and (iv) if $\mathcal{G}^\prime$ is uniform at the horizon, the skin depth vanishes on the whole of the horizon. The vanishing of the skin depth (associated with the magnetic field penetration) in this manner suggests that the plasma becomes a perfect superconductor wherever $\partial\mathcal{G}^\prime/\partial\theta=0$ on the horizon (which implies $\vec{B}=0$ on the horizon), i.e, the vanishing of the GGV is entirely equivalent to the vanishing of the magnetic field. In contrast to the expulsion of magnetic fields for extremal black holes (as described in \cite{Kingetal1975,BicakDvorak1980,Chamblinetal1998,bivcak2015near,gurlebeck2017meissner,gurlebeck2018meissner,kunz2017magnetized}), the vanishing of the skin depth $\lambda$ for the plasma states we present here does not require extremal or near extremal spin; it works even for non-spinning Schwarzschild black holes. Our result therefore demonstrates what could have been expected - that the geometry near stationary black hole horizons can have a significant effect on the electrodynamics  of the surrounding plasma. The next step in the investigation of magnetized plasmas surrounding black holes should consider a broader class of velocity profiles and more general (non-superconducting) equilibrium states; this will be discussed in future work.

\begin{acknowledgments}
S. M. M.'s work has been supported by US DOE Contract No. DE-FG02-04ER-54742.
\end{acknowledgments}

\bibliography{ref_new} 

\end{document}